\documentstyle[preprint,aps]{revtex}
\begin{document}
\def \sign{ \mathop{ \rm sign}\nolimits}
\def \trace{ \mathop{ \rm trace}\nolimits}
\draft
\title{Integrable quantum chains combining site states in different
representations of $su(3)$}
\author{J. Abad and M. Rios}
\address{Departamento de F\'{\i}sica Te\'{o}rica, Facultad de Ciencias\\
Universidad de Zaragoza, 50009 Zaragoza, Spain}
\date{\today}
\maketitle
\begin{abstract}
The general expression for the local matrix $L(\theta)$ of a quantum chain with the site space in any representation of  $su(3)$ is
obtained.
 This is made by generalizing $L(\theta)$ from the fundamental
representation and imposing the fulfilment of the Yang-Baxter equation.
Then, a non-homogeneous spin chain combining different representations
of $su(3)$ is solved by a method inspired in the nested Bethe ansatz.
The solution for the eigenvalues of the trace of the monodromy matrix  is given
as two coupled Bethe equations. A conjecture about the solution of a chain with
the site states in different representations of $su(n)$ is presented
\end{abstract}
We present the study of solvable alternating chains whose site states of which
are a
mixture of any two representations of $su(3)$. We made an approach to this
problem in  our paper \cite{ri},
where we solved an alternating chain mixture of the two fundamental
representations of  $su(3)$ and presented a method, a modification of the
nested
Bethe ansatz (MNBA), needed to find the  Bethe equation (BE)
solutions of the problem.

A chain is defined by giving the monodromy matrix that, for a
non-homogeneus chain, we call  $T^{\rm{alt}}$. This matrix is a product of local matrices,
 generally designed by $L(\theta)$, that are defined in an
auxiliary space and a site space \cite{rni}. The auxiliary space will be always the
$\{3\}$ representation of $su(3)$, whereas for the site space we will take any
representation of the same algebra.

We denote a representation of $su(3)$ by the indices of its associated Dynkin
diagram
$(m_1, m_2)$, where $m_1$ and $m_2$ correspond to the $\{3\}$ and
$\{\bar{3}\}$ representations respectively. In the figures a continuous
line is used for the fundamental representation $(1,0)$ and a wavy line for
any other
representation. Thus, the operators $L(\theta)$  are denoted as indicated in
figure 1. Besides, in order to simplify the writing of the formulae, we will
adopt the following identifications: $L(\theta) \equiv L^{(1,0)(1,0)}(\theta)$
and
${L'}(\theta) \equiv L^{(1,0)(m_{1},m_{2})}(\theta)$.

\bigskip
\input epsf
\centerline{\epsfxsize=10cm  \epsfbox{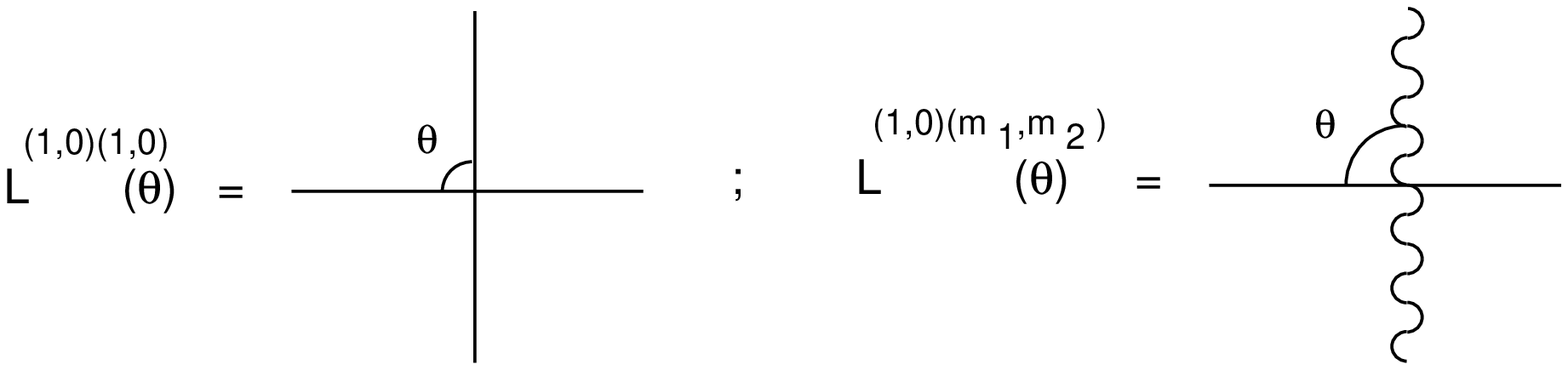}}
\centerline{Figure. 1}
\bigskip
The operator $L(\theta)$ can be written \cite{ri}

\begin{equation}\label{ei}
L(\theta)=
\pmatrix{
{1 \over 2}(\lambda^{3}q^{-N^{\alpha}}-\lambda^{-3}q^{N^{\alpha}}) & {1 \over 2}\lambda 
{(q^{-1}-q)
} f_{1} & {1 \over 2}\lambda^{-1}{(q^{-1}-q)} [f_{2},f_{1}] \cr
{1 \over 2}\lambda^{-1}{(q^{-1}-q) } e_{1} &{1 \over 2}(\lambda^{3}q^{-N^{\beta}}-
\lambda^{-3}q^{N^{\beta}}) & {1 \over 2} \lambda {(q^{-1}-q)} f_{2} \cr
{1 \over 2} \lambda{(q^{-1}-q)} [e_{1},e_{2}]  & {1 \over 2}\lambda^{-1}{(q^{-1}-q)}
e_{2} &
{1 \over 2}(\lambda^{3}q^{-N^{\gamma}}-\lambda^{-3}q^{N^{\gamma}}) \cr
},
\end{equation}

where the parameters  $\lambda$ and $q$ have been taken as the functions of
$\theta$ and $\gamma$
\begin{equation}\label{eii}
\lambda = e^{\theta \over 2}, \qquad q = e^{-\gamma},
\end{equation}

the $N$ matrices are
\begin{mathletters}
\label{eai}
\begin{eqnarray}
N^{\alpha}&&={2 \over 3}h_{1}+{1 \over 3}h_{2}+{1 \over 3}I, \label{eaia}\\
\cr
N^{\beta}&&=-{1 \over 3}h_{1}+{1 \over 3}h_{2}+{1 \over 3}I, \label{eaib}\\
N^{\gamma}&&=-{1 \over 3}h_{1}-{2 \over 3}h_{2}+{1 \over 3}I, \label{eaic}
\end{eqnarray}
\end{mathletters}
and $\{e_{i}, f_{i}, q^{\pm h_{i}}\}$, $i=1, 2$ are the Cartan generators of the
deformed algebra $U_{q}(sl(3))$.

To obtain the operators ${L'}(\lambda) $ with the new parameters given in
(\ref{eii}),
we take (\ref{ei}) as a basis and write
\begin{equation}
\label{eiv}
L'(\lambda)=
\pmatrix{
{1 \over 2}(\lambda^{3}q^{-N^{\alpha}}-\lambda^{-3}q^{N^{\alpha}}) & \lambda
 F_{1} & \lambda^{-1} F_{3} \cr
\lambda^{-1} E_{1} &{1 \over 2}(\lambda^{3}q^{-N^{\beta}}-
\lambda^{-3}q^{N^{\beta}}) & \lambda  F_{2} \cr
\lambda E_{3}  & \lambda^{-1} E_{2} &
{1 \over 2}(\lambda^{3}q^{-N^{\gamma}}-\lambda^{-3}q^{N^{\gamma}}) \cr
},
\end{equation}
where the operators $\{E_{i},F_{i}\}$, $i=1,2,3,$ are  unknown and will be
determined by imposing the Yan Baxter equation (YBE)
\begin{equation}\label{ev}
R(\lambda/\mu)[L'(\lambda) \otimes L'(\mu)] =
[L'(\mu) \otimes L'(\lambda)]R(\lambda/\mu)
\end{equation}
\bigskip
\centerline{\epsfxsize=10cm  \epsfbox{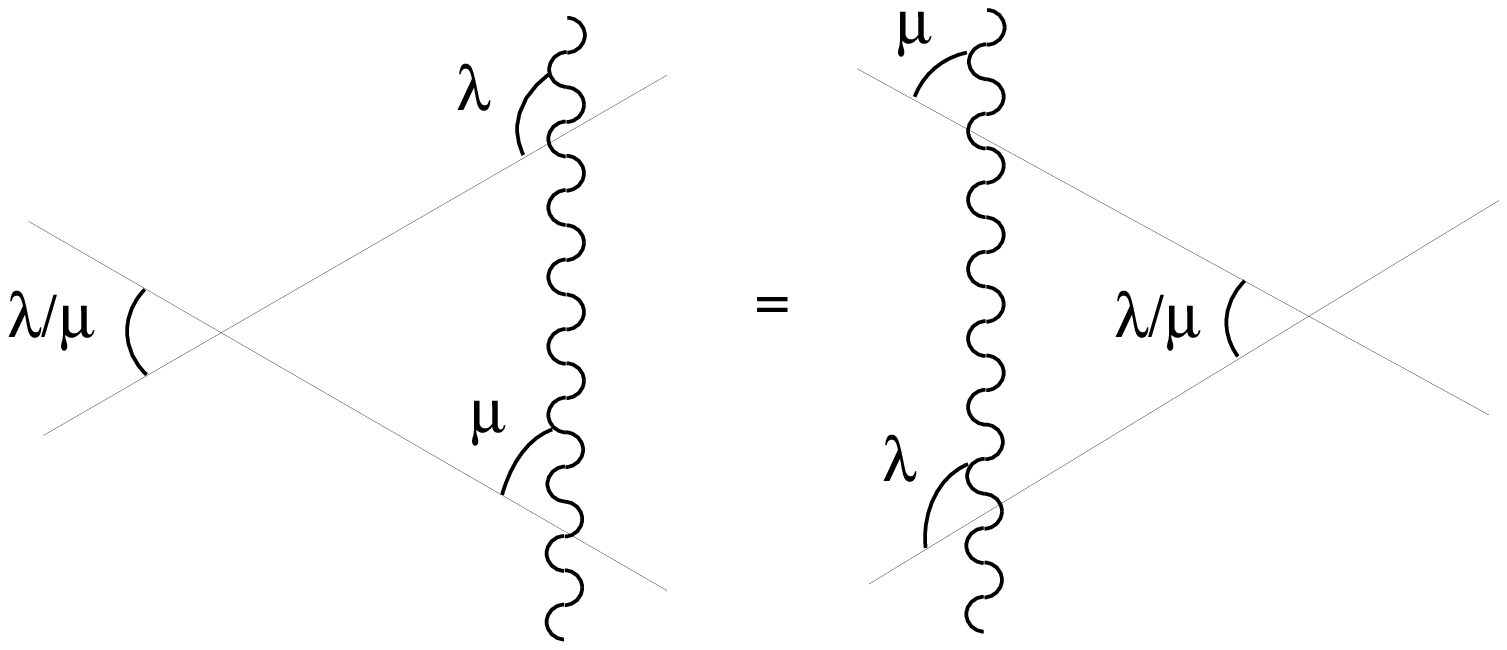}}
\centerline{Fig. 2}
\bigskip
\noindent shown in figure 2.
The elements $R_{c, a} ^ {b,d}(\theta)\equiv [ L_{a,b} (\theta)]_{c,d}$ of $R(\lambda/\mu)$ 
are given
\cite{rii} by
\begin{equation}\label{evi}
R(\lambda,\mu)=\left (\matrix{
a & 0 & 0 & 0 & 0 & 0 & 0 & 0 & 0 \cr
0 & d & 0 & b & 0 & 0 & 0 & 0 & 0 \cr
0 & 0 & c & 0 & 0 & 0 & b & 0 & 0 \cr
0 & b & 0 & c & 0 & 0 & 0 & 0 & 0 \cr
0 & 0 & 0 & 0 & a & 0 & 0 & 0 & 0 \cr
0 & 0 & 0 & 0 & 0 & d & 0 & b & 0 \cr
0 & 0 & b & 0 & 0 & 0 & d & 0 & 0 \cr
0 & 0 & 0 & 0 & 0 & b & 0 & c & 0 \cr
0 & 0 & 0 & 0 & 0 & 0 & 0 & 0 & a \cr
}\right ),
\end{equation}
with
\begin{mathletters}
\label{eri}
\begin{eqnarray}
a(\lambda,\mu)&={1 \over 2}(\lambda^{3}\mu^{-3}q^{-1} -
\lambda^{-3}\mu^{3}q) , \label{eria} \\
b(\lambda,\mu)&={1 \over 2}(\lambda^{3}\mu^{-3} -
\lambda^{-3}\mu^{3}) , \label{erib} \\
c(\lambda,\mu)&={1 \over 2}(q^{-1}-q)\lambda\mu^{-1} , \label{eric} \\
d(\lambda,\mu)&={1 \over 2}(q^{-1}-q)\lambda^{-1}\mu . \label{erid}
\end{eqnarray}
\end{mathletters}

The relations obtained are
\begin{mathletters}
\label{eaii}
\begin{eqnarray}
E_{1}q^{N^{\alpha}} &&= q^{-1}q^{N^{\alpha}} E_{1}, \label{eaiia} \\
E_{1}q^{N^{\beta}} &&= qq^{N^{\beta}} E_{1}, \label{eaiib} \\
F_{1}q^{N^{\alpha}} &&= qq^{N^{\alpha}} F_{1}, \label{eaiic} \\
F_{1}q^{N^{\beta}} &&= q^{-1} q^{N^{\beta}} F_{1}, \label{eaiid} \\
E_{2}q^{N^{\alpha}} &&= qq^{N^{\alpha}} E_{2}, \label{eaiie} \\
E_{2}q^{N^{\beta}} &&= q^{-1}q^{N^{\beta}} E_{2}, \label{eaiif} \\
F_{2}q^{N^{\alpha}} &&= q^{-1}q^{N^{\alpha}} F_{2},\label{eaiig} \\
F_{2}q^{N^{\beta}} &&= qq^{N^{\beta}} F_{2} ,\label{eaiih} \\
\left[ E_{1},F_{1} \right]&&=(q^{-1}-q)\left(q^{N^{\beta}-N^{\alpha}} -
q^{N^{\alpha}-N^{\beta}} \right), \label{eaiiil} \\
\left[ E_{2},F_{2} \right]&&=(q^{-1}-q)\left(q^{N^{\gamma}-N^{\beta}} -
q^{N^{\beta}-N^{\gamma}} \right), \label{eaiij} \\
E_{3} &&= {1 \over (q^{-1}-q)}q^{-N^{\beta}}[E_{1},E_{2}] ,\label{eaiik} \\
F_{3} &&= {1 \over (q^{-1}-q)}q^{N^{\beta}}[F_{2},F_{1}] .\label{eaiil}
\end{eqnarray}
\end{mathletters}
Besides, the modified Serre relations
\begin{mathletters}
\label{eaiii}
\begin{eqnarray}
&&q^{-1}E_{1}E_{1}E_{2}-(q+q^{-1})E_{1}E_{2}E_{1} +
qE_{2}E_{1}E_{1} =0, \label{eaiiia} \\
&&qE_{2}E_{2}E_{1}-(q+q^{-1})E_{2}E_{1}E_{2} +
q^{-1}E_{1}E_{2}E_{2} = 0, \label{eaiiib} \\
&&q^{-1}F_{1}F_{1}F_{2}-(q+q^{-1})F_{1}F_{2}F_{1} +
qF_{2}F_{1}F_{1} = 0, \label{eaiiic} \\
&&qF_{2}F_{2}F_{1}-(q+q^{-1})F_{2}F_{1}F_{2} +
q^{-1}F_{1}F_{2}F_{2} = 0 ,\label{eaiiid}
\end{eqnarray}
\end{mathletters}
should be verified.

It must be noted that that the relations (\ref{eaii}) are the usual ones for
the quantum group
$U_{q}(sl(3))$ while the relations  (\ref{eaiii}) are not the usual ones. Since the generators $e_i$ and $f_i$ of the deformed algebra do not satisfy the YBE, we must introduce the operators $E_i$
and $F_i$ defined by
\begin{mathletters}
\label{eaiv}
\begin{eqnarray}
F_{i}&&={1 \over 2}(q^{-1}-q)Z_{i}f_{i},  \label{eaiva} \\
E_{i}&&={1 \over 2}(q^{-1}-q)e_{i}Z_{i}^{-1} ,\qquad \qquad i=1,2
\label{eaivb}
\end{eqnarray}
\end{mathletters} 
where $e_i$ and $f_i$, $i=1,2,$ are the generators of $U_{q}(sl(3))$ in the
representation $(m_1 , m_2)$ and $Z_i$ are two diagonal operators that can 
be obtained by imposing the verification of the relations (\ref{eaii}) and
(\ref{eaiii}). In
this way, one obtains the general form of these operators given by
\begin{mathletters}
\label{eav}
\begin{eqnarray}
Z_{1}&&=q^{a_{1}h_{1}-{1 \over 3}h_{2}+a_{3}I} , \label{eava} \\
Z_{2}&&=q^{{1 \over 3}h_{1}+(a_{1}+{1 \over 3})h_{2}+b_{3}I} , \label{eavb}
\end{eqnarray}
\end{mathletters}  
where the operators $h_i$, $i=1,2,$ are the diagonal elements of the algebra
$sl(3)$, and $a_1, a_3$ and $ b_3$ are free parameters associated with
the transformations that leave the YBE invariant.

The knowledge of the operator $L' $ permits us to build a family of solvable
multistate  chains, based on the $su(3)$ algebra, that mixes various
representations. The monodromy operator
 is built as a product of local operators. For instance, for a chain 
which alternates the representations $(1, 0)$ and $(m_1, m_2)$,
\begin{equation}\label{evii}
T^{\rm (alt)}_{a,b} (\theta)= L^{(1)}_{a,a_1}(\theta) {L' }
^{(2)}_{a_1,a_2}
(\theta) \ldots
L^{(2N-1)}_{a_{2N-2},a_{2N-1}}(\theta){L' }^{(2N)}_{a_{2N-1,b}}
(\theta) ,
\end{equation}
that can be represented graphically as shown in figure 3.

\bigskip
\centerline{\epsfxsize=10cm  \epsfbox{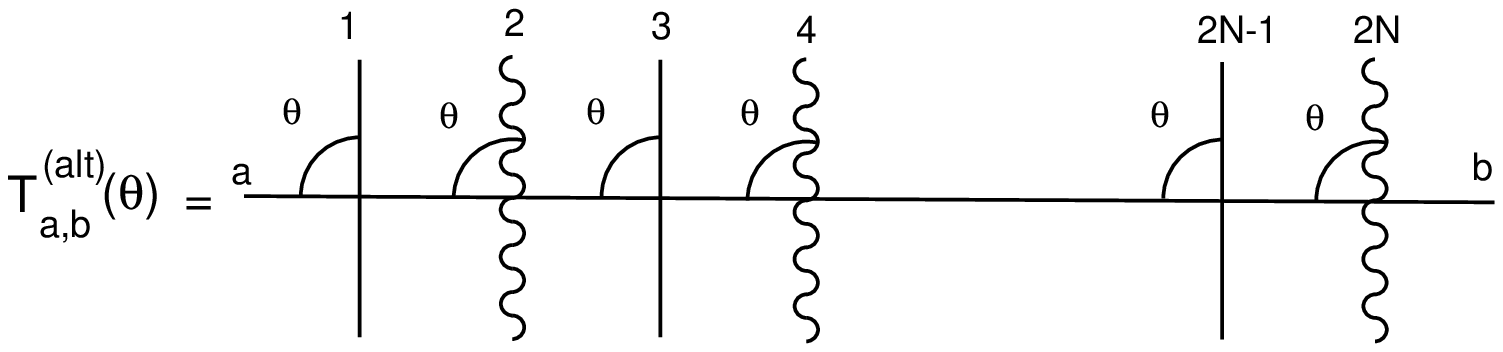}}
\centerline{Figure 3}
\bigskip

The monodromy operator can be written as a matrix in the auxiliary space whose
elements are operators in the state space of the chain
\begin{equation}\label{eviii}
T^{\rm{alt}}(\theta)=
\pmatrix{
A(\theta) & B_2 (\theta) & B_3 (\theta) \cr
C_2 (\theta)& D_{2,2}(\theta) & D_{2,3}(\theta) \cr
C_3 (\theta)& D_{3,2}(\theta) & D_{3,3}(\theta) \cr
}.
\end{equation}
We are interested in finding the eigenvalues of the trace of $T$.

Due to the fact that in this type of systems one cannot find a pseudovacuum eigenstate 
of the diagonal operators $A$ and $D_{i,i}$ and such that the action on it of the
non-diagonal $D_{i,j}$ be null, we must use a modification of the nested Bethe
ansatz (MNBA) to find the Bethe equations solutions of the problem
\cite{riii}.

The method in this case has two steps as the usual NBA. In the first step, we
start by building a subespace in the space of states whose elements are
eigenstates of $A$ and stable under the action of the $D_{i,j}$ operator. A
state of this subspace is taken as pseudovacuum state.  The general state is
obtained by applying  $r$ operators $B_{i}(\mu_i), i=1,...,r,$  on this state, as
in the habitual NBA method. In the second step we must repeat the first step
with a two by two monodromy matrix $T^{(2)}$ that is the product of the $D$
submatrix of $T^{(\rm{alt})}$ and a matrix that relates two products of  $r$
operators $B_{i}(\mu_i)$ differing in a cyclic permutation.  The usual NBA
can be only applied when $T^{(2)}$ is solely  product of cyclic permutation
operators . This new matrix verifies a Yang-Baxter equation and we can repeat
the first step, finishing the process. The method is described in ref.
\cite{ri} where it is applied to  non-homogeneus chain that mixes the $\{3\}$ and
$\{3^*\}$ representations.

The results depend on the representation of $su(3)$ in which the states of
every site are. If the site space is in the representation $(m_1, m_2)$, the highest
weight state is eigenstate of the diagonal operators in (\ref{eiv}), $L'_{i,i}
(\theta)=L^{(1,0),(m_1, m_2)}_{i,i} (\theta), \quad i=1,2,3,$  with
eigenvalues
\begin{mathletters}
\label{eaix}
\begin{eqnarray}
l^{m_1, m_2}_{1,1}(\theta) & = & \sinh (\frac{3}{2} \theta +
(\frac{2}{3}m_{1} + \frac{1}{3}m_{2} + \frac{1}{3})\gamma),
\label{eaixa}  \\
l^{m_1, m_2}_{2,2}(\theta) & = & \sinh (\frac{3}{2} \theta +
(-\frac{1}{3}m_{1} + \frac{1}{3}m_{2} + \frac{1}{3})\gamma),
\label{eaixb}  \\
l^{m_1, m_2}_{3,3}(\theta) & = & \sinh (\frac{3}{2} \theta +
(-\frac{1}{3}m_{1} - \frac{2}{3}m_{2} + \frac{1}{3})\gamma),
\label{eaixc}
\end{eqnarray}
\end{mathletters}
Then, each site  introduces the  source functions
\begin{mathletters}
\label{eax}
\begin{eqnarray}
g_{1}(\theta) &&={l^{m_1, m_2}_{1,1} (\theta)\over l^{m_1, m_2}_{2,2}(\theta)
}={\sinh(
{3 \over 2}\theta+({2 \over 3}m_{1}+{1 \over 3}m_{2}+
{1 \over 3})\gamma) \over \sinh
({3 \over 2}\theta+(-{1 \over 3}m_{1}+{1 \over 3}m_{2}+
{1 \over 3})\gamma)}, \label{eaxa} \\
g_{2}(\theta) &&={ l^{m_1, m_2}_{3,3}(\theta)\over l^{m_1, m_2}_{2,2}(\theta)
}={\sinh
({3 \over 2}\theta+(-{1 \over 3}m_{1}-{2 \over 3}m_{2}+
{1 \over 3})\gamma) \over \sinh
({3 \over 2}\theta+(-{1 \over 3}m_{1}+{1 \over 3}m_{2}+
{1 \over 3})\gamma)}. \label{eaxb}
\end{eqnarray}
\end{mathletters}

The Bethe equations for a general chain combining $N_1$ sites in the $(m_1,
m_2)$ representation and $N_2$ sites in the $(m'_1, m'_2)$ representation with
source functions $g^{(1)}_i$ and  $g^{(2)}_i ,\quad i=1,2,$  respectively, are

\begin{mathletters}
\label{eaxi}
\begin{eqnarray}
[g^{(1)}_{1}(\mu_{k})]^{{N}_1}[g^{(2)}_{1}(\mu_{k})]^{{N}_2}
&&=\prod_{j=1 \atop j \neq
k}^{r} {g(\mu_{k}-\mu_{j}) \over g(\mu_{j}-\mu_{k})}\prod_{i=1}^
{s}g(\lambda_{i}-\mu_{k}), \quad k=1\cdots r , \label{eaxia} \\
\left[ g^{(1)}_{2}(\lambda_{k})
\right]^{{N}_1}[g^{(2)}_{2}(\lambda_{k})]^{{N}_2}&&=
\prod_{j=1}^{r}g(\lambda_{k}-\mu_{j})
\prod_{i=1 \atop i \neq
k}^{s} {g(\lambda_{i}-\lambda_{k}) \over g(\lambda_{k}-\lambda_{i})},
\quad k=1\cdots s
, \label{eaxib}
\end{eqnarray}
\end{mathletters}
where $g$ is the $g_1$ source function of the fundamental representation $(1,
0)$
\begin{equation}\label{eaxii}
g(\theta)={\sinh({3 \over 2}\theta + \gamma) \over \sinh({3 \over
2}\theta)}.
\end{equation}

The eigenvalue of the trace of the monodromy matrix is

\begin{eqnarray}
	 \lefteqn{\Lambda(\theta) = [l^{(1)}_{1,1}(\theta)]^{N_{1}}
	 [l^{(2)}_{1,1}(\theta)]^{N_{2}} \prod_{j=1}^{r}
	 g(\mu_{j}-\theta) +
	 }
	\nonumber  \\
	 &  & + \prod_{j=1}^{r} g(\theta-\mu_{j}) \left\{
	 [l^{(1)}_{2,2}(\theta)]^{N_{1}}
	 [l^{(2)}_{2,2}(\theta)]^{N_{2}}\prod_{i=1}^{s}
	 g(\lambda_{i}-\theta) +
	 \right.
	\nonumber  \\
	 &  & + \left.
	 [l^{(1)}_{3,3}(\theta)]^{N_{1}}
	 [l^{(2)}_{3,3}(\theta)]^{N_{2}}
	 \prod_{l=1}^{r}\frac{1}{g(\theta-\mu_{l})} \prod_{i=1}^{s}
	 g(\theta-\lambda_{i})
	 \right\}.
	\label{exiii}
\end{eqnarray}

In the light of this, the generalization to the case of
mixed chains with more than two different representations seems simple, although
the physical
models that they represent will be less local and the interaction more complex.

In a non-homogeneous chain  combining different representations of $su(n)$,
each  representation introduces $(n-1)$
functions ( that we call source functions). Each solution will have $(n-1)$
sets of equations (the same number of dots in its Dynkin diagram). The left 
hand side of the equations will be a product of the respective source functions
powered to the number of sites of each representation and the right hand side 
a productof source functions similar to (\ref{eax}).

\end{document}